%% file: main.tex
\documentclass[12pt]{article}

\usepackage{
    amsmath,
    amsthm,
    amssymb,
    natbib,
    verbatim,
    upref,
    paralist,
    indentfirst,
    verbatim,
    graphicx,
    pdfpages,
    caption,
    hyperref,
    nicefrac,
    enumitem,
    array,
    subcaption,
    tikz,
    multirow,
    booktabs,
    placeins
}
\usetikzlibrary{
arrows,
positioning,
shapes.geometric
}

\theoremstyle{definition}

\usepackage[T1]{fontenc}
\usepackage[utf8]{inputenc}
\usepackage{authblk}
\usepackage{tgheros}
\usepackage[margin=1in]{geometry}
\usepackage{setspace}
\begin{document}

\title{Artificial Intelligence for Multi-Unit Auction design\thanks{This research is supported by the Translation Investment Fund (TIF23015), RMIT University. We would like to acknowledge various seminar participants for their insightful comments. The usual disclaimer applies. }
}

\author{Peyman Khezr\thanks{School of Economics, Finance and Marketing, RMIT University, Melbourne, Australia. Corresponding author, email: peyman.khezr@rmit.edu.au.} \quad \quad Kendall Taylor\thanks{School of Computing Technology, RMIT University, Melbourne, Australia.}}

\date{}
\maketitle

\begin{abstract}
Understanding bidding behavior in multi-unit auctions remains an ongoing challenge for researchers. Despite their widespread use, theoretical insights into the bidding behavior, revenue ranking, and efficiency of commonly used multi-unit auctions are limited. This paper utilizes artificial intelligence, specifically model-free reinforcement learning, to simulate bidding in three prominent multi-unit auctions employed in practice. We introduce six algorithms that are suitable for learning and bidding in multi-unit auctions and compare them using an illustrative example. This paper underscores the significance of using artificial intelligence in auction design, particularly in enhancing the design of multi-unit auctions.
\end{abstract}
\noindent

\vspace{1ex}


\section{Introduction}

Multi-unit auctions play a fundamental role in the allocation of goods and services across various markets, including treasury bills, emission permits, spectrum licenses, and electricity \citep{Khezr2022, bichler2017handbook}. These auctions are often characterized by asymmetric information and, due to the existence of multiple units, result in a complex bidding environment. Although multi-unit auctions are commonly used in practice, there is a lack of consensus in the literature regarding their performance in terms of revenue and efficiency. Moreover, economic theory often lacks clear guidance on the outcomes of many multi-unit auctions \citep{Krishna2009}.

A range of disciplines, including economics and computer science, have applied various methodologies to investigate multi-unit auctions. Auction theory, data analytics, experimental approaches, and simulations represent some of the predominant methods used in these studies \citep{ausubel2014demand, roughgarden2017price, morgenstern2016learning, hailu2007designing}. Each method offers distinct advantages but also comes with limitations. A common flaw in these approaches is their inability to comparatively assess the performance of different auction variations under similar conditions. Additionally, traditional simulation methods often lack clear guidance as they require an existing model or theoretical prediction to feed into the simulation. In this article, we use artificial intelligence, specifically introducing several reinforcement learning (RL) algorithms, to simulate bidding behavior in various multi-unit auctions.

Reinforcement learning (RL) has been recognized as an effective means of simulating learning processes in uncertain environments, such as auctions \citep{silver2021reward}. The capability of RL-based algorithms to learn in a model-free manner makes them particularly well-suited to contexts like multi-unit auctions, where predictive models are often unclear or unavailable. This paper concentrates on three primary RL approaches: Q-learning, Policy Gradient, and Actor-Critic. We discuss six algorithms that are suitable for simulating the learning and bidding processes in auctions. Our findings detail the principal advantages and disadvantages of each algorithm in the context of bidding behavior simulation. This paper represents one of the initial attempts to apply artificial intelligence to analyze multi-unit auctions.

Our approach leverages artificial intelligence as a complementary tool to previous methodologies, aiming to emulate the human thought process and decision-making in complex bidding scenarios within multi-unit auctions. Using a standard private value model, we assume that buyers' valuations for each unit up for auction are privately known. Our investigation focuses on three sealed-bid multi-unit auctions: the Discriminatory Price (DP), the Generalized Second-Price (GSP), and the Uniform-Price (UP) auctions. All three of these auctions have been extensively used in real-world markets to sell various goods and services.

We first provide a metric for learning and demonstrate through an example, in which each bidder demands two units, that almost all RL algorithms perform very well in terms of learning to bid. Although they vary in terms of speed and learning patterns, some top-performing algorithms surprisingly converge to very similar strategies in the final episodes of learning. One of the six algorithms, namely Proximal Policy Optimization, not only performs the best by earning the highest payoffs in all cases, but is also the most stable in terms of learning. We then compare the revenue and efficiency of the three auctions in three different scenarios.

Our results indicate that the uniform-price auction performs very well in terms of efficiency, followed by the generalized second-price auction. The discriminatory price auction, however, dominates in terms of revenue in most cases, except when the number of units available is the lowest.

\subsection*{Background}

Since the seminal work by Vickrey \citep{vickrey1961counterspeculation}, it is known that bidding behavior in multi-unit auctions does not necessarily follow that in single-unit auctions. Despite this, most conventionally used multi-unit auctions are extensions of a standard single-unit format. For instance, the generalized second-price auction, as appears from its name, is a format that extends the second-price rule to a multi-unit case. There are several studies that show these multi-unit extensions do not necessarily have the ideal properties of the single unit format \citep[e.g.][]{back1993auctions, edelman2007internet, ausubel2014demand}. Additionally, due to multiple equilibria and lack of closed-form solutions for the bidding functions, theoretical studies have no clear conclusion about the revenue and efficiency ranking of commonly used multi-unit auctions \citep{Krishna2009, ausubel2014demand}.

Recently, the use of RL approaches, particularly Q-learning algorithms, has become more prevalent in the auction theory literature. For example, \cite{banchio2022artificial} employ a simple Q-learning algorithm to simulate bidding behavior in two different single-unit auction formats. \cite{khezr2024strategic} is another study that uses Q-learning to simulate a more complex type of auction, namely \textit{the knapsack auction}. While various other RL approaches could potentially be used to simulate bidding in auctions, current studies are mainly limited to simple Q-learning algorithms. Therefore, our main aim is to explore other, more advanced algorithms that could effectively simulate such complex bidding environments.


\section{Model}

A seller owns $K>2$ units of a homogeneous object. There are $n>1$ potential buyers, each demanding up to $k>1$ units of the object. Every buyer $i$ has a distinct marginal value, $v^i_k>0$, for the $k$th unit of the object. We use $\mathbf{v}^i= (v_1^i, v_2^i, ..., v_k^i)$ to denote the vector of values for buyer $i$ with diminishing marginal values, that is, $ v_1^i\geq v_2^i \geq ... \geq v_k^i$. Each buyer privately knows these values. However, it is publicly known that these values are independently and identically distributed according to a distribution function $F_i(.)$ with finite bounds. Finally, to avoid trivial cases, we assume that the sum of the units demanded by all buyers exceeds the number of available units, that is, $\sum_i k_i > K$.

\subsection*{The allocation mechanism}

The seller uses a sealed-bid multi-unit auction to allocate the objects. Each bidder submits a vector of bids $\mathbf{b}^i$ for all units they demand. The auctioneer ranks the bids from the highest to the lowest and allocates the object to the $K$ highest submitted bids. We explore three types of payment rules as follows.

\subsection*{Discriminatory Price Auction (DP)}
In the discriminatory price auction, also known as pay as bid auction, every buyer $i$ submits up to $k$ bids, represented by $\mathbf{b}^1 = (b_1^i, b_2^i,..., b_k^i)$. The auctioneer arranges all the $n\times k$ bids from the highest to the lowest. The top $k$ bids each win a unit and pay the amount they bid for that particular unit. For example, if the first bid from bidder three, $b_1^3$, holds the top rank, she secures the first unit and pays $b_1^3$ for it.

\subsection*{Generalized Second-Price Auction (GSP)}

Similarly to the DP auction, each buyer $i$ submits up to $k$ bids. The top $k$ bids each win a unit. However, in this case, winners pay the amount of the bid that is ranked immediately below their own (excluding their own subsequent bids). For instance, if bidder three's first bid, $b_1^3$, holds the highest rank, she secures the first unit but pays the amount of the second-highest bid. 

\subsection*{Uniform-Price Auction (UP)}
The Uniform Price Auction, as one of the most popular auctions in practice, aims to replicate competitive equilibrium pricing by setting a uniform price for all units won in the auction. Similar to the previous two auctions, the top $k$ bids each win a unit. However, in this auction, every winning bidder pays a uniform price for each unit won, which is equal to the highest losing bid---that is, the $K+1$th highest bid.







\section{Reinforcement Learning approaches}

We use Reinforcement Learning (RL) methods to model bidder behavior in multi-unit auctions with multiple bidders. Inspired by behavioral psychology, RL algorithms guide an \emph{agent} to learn how to accomplish a goal via feedback (consisting of rewards and penalties). An agent, in the context of this work, is a simulated human bidder operating within an \emph{environment}, an auction, in which they interact with other agents via bidding for available items. The environment encompasses everything the agent interacts with and can change over time, either due to the agent's actions or independently. Interactions between an agent and the environment are defined by the terms \emph{state}, \emph{action}, and \emph{reward}.

A \emph{state} is a complete description of the agent's situation within the environment. An \emph{observation}, which might be a partial view of the state, provides the agent with information about the current situation. \emph{Actions} are the set of all possible decisions the agent can make. A \emph{reward} is a feedback signal indicating the direct effect of the agent's previous action. The agent's objective is to maximize their return (or cumulative reward), over time. This objective and the associated decision-making strategy are encapsulated in an agent's \emph{policy}, a mapping of states (or observations) to actions.

The RL problem is fundamentally about learning the optimal policy that maximizes the expected return when starting from any given state. The agent gradually improves its decision-making abilities by exploring and exploiting different actions. These methods are adapted to the complexity and uniqueness of each auction scenario presented in this paper. Our approach involves multiple agents (bidders) in a dynamic, discrete environment. We evaluate three main approaches for this study:

\begin{enumerate}
    \item \textbf{Q-Learning:}\\
    Traditional tabular and deep learning approaches are assessed, where an agent learns from their own actions, states, and rewards. In the tabular version, each agent has a Q-table (developed over many auction episodes), which guides them to the best action for a given state that maximises rewards. Initially, exploration is emphasised, but over time, exploiting known actions becomes dominant. The deep Q-learning version (DQN) uses a neural network instead of a Q-table. This improves the handling of large state-action spaces and enables better generalisation in complex environments.
    
    \item \textbf{Policy Gradient:}\\
    Again, traditional and deep learning versions are considered. The policy gradient approach models action probabilities for given states, aiming to maximise rewards. The policy is parameterised and updated based on the reward gradient.
    A neural network represents the policy using deep learning, enhancing generalisation and scalability and allowing for more complex policy learning.
    
    \item \textbf{Actor-Critic:}\\
    This method combines value-based estimates (in ways similar to Q-learning) and policy-based functions (similar to those used in Policy Gradient algorithms). The \emph{actor} proposes actions, while the \emph{critic} evaluates them and guides policy updates using neural networks. Two popular and successful methods are evaluated: the Advantage Actor-Critic (A2C) and Proximal Policy Optimization (PPO) algorithms. These methods differ in their approaches to estimating the value of actions and updating the policy.
\end{enumerate}

\subsection{Environment setup and evaluation}
A typical RL problem trains an agent using multiple independent sets of agent and environment interactions called an \emph{episode}. An episode is a complete sequence of states, actions, and rewards that ends with a terminal state. It consists of a series of steps, where at each step, the agent receives an observation of the environment's current state, takes an action based on that observation, and receives a reward from the environment. For example, in a chess game, an episode would be one complete game from the initial board setup until checkmate, resignation, or a draw is reached. The agent and environment, therefore, interact in a loop:
\begin{enumerate}
    \item The environment presents a state to the agent
    \item The agent takes an action based on the state
    \item The environment transitions to a new state and provides a reward to the agent
    \item The process repeats from step 1 (until the termination condition is met)
\end{enumerate}

In this paper's multi-agent, multi-unit auction scenario, an episode includes only a single step. The number of bids an agent can submit is the same for all agents for all episodes in a training or evaluation session. The state is unique for each agent (but may be identical) and represents the agent's valuation of each item on offer. Each agent's item valuations are drawn from a random distribution (with replacement), for each episode. An episode's state is independent of all other episodes and, possibly, may not be repeated during a session. The single-step episodes differ from typical RL situations and can confound the learning process further.

The reward function used for all auction types is as follows:
\begin{equation}\label{eqrew}
   R = 
\begin{cases}
    \nicefrac{p}{max(v,~1.0)},      & \text{if } s = 1 \text{ and } p > 0\\
    \nicefrac{-(v-p)}{max(v,~1.0)},  & \text{if } s = 1 \text{ and } p \leq 0\\
    -0.01,                          & \text{if } s = 0,
\end{cases} 
\end{equation}
where $R$ is the reward function, $s$ is a binary variable representing whether the bid is successful, $p$ is the agent's payoff given their bid, and $v$ is the item value associated with a particular bid.

The reward function, therefore, returns values that can be either positive or negative, rewarding positive payoffs and punishing overbidding more severely than unsuccessful bidding. Specifically, when $p>0$, the agent receives a positive reward as determined by Equation \ref{eqrew}. However, when $p<0$, it indicates that the agent's payment—whether it is their own bid (in DP) or some other lower bid (in GSP and UP)—exceeds their value thus the numerator of the reward function in Equation \ref{eqrew} becomes negative. In this case, the agent receives a penalty in the form of a negative reward. Both reward and punishment are proportional to the value, which means that for higher values, any given positive or negative payoff is less rewarded or punished respectively. For instance, if an agent’s payoff is 2, it is more significantly rewarded when their value is 3 as opposed to when their value is 9. This design ensures that the utility function remains concave for all agents.



Next we outline six RL algorithms used in the auction simulations: two Q-learning methods, two Policy Gradient, and two Actor-Critic approaches. Each algorithm is implemented as a \emph{bidder} who places a bid in relation to the utility or \emph{value} they place on each of the items on offer. The algorithms are all established and proven methods, each possessing unique strengths and weaknesses (see Table \ref{tbl:Comparison}).

\subsection{Q-Learning}
Q-Learning is an off-policy model-free RL method developed by \cite{watkins1989learning}. The method learns the value of an action in a given state without requiring a model of the environment. A key device used by Q-Learning is a data store called a Q-table. The Q-table records expected rewards for all possible actions in a given state. The Q-table is updated using a formulation developed by \cite{bellman1957dynamic}. The Bellman equation considers several factors, including the current state, the reward obtained from the action taken, the maximum expected reward for the new state, and a discount rate that determines the importance of future rewards. This process is performed iteratively to help the agent learn the optimal policy by choosing actions with the highest expected rewards. 

For each iteration, the Q-value for a state-action combination $Q(s, a)$ is updated using the Bellman equation:
\begin{equation}
Q^{new}(s,a) \leftarrow Q(s,a) + \alpha (r + \gamma \max_a\left(Q(s',a') - Q(s,a)\right),
\end{equation}
where $Q^{new}(s,a)$ is the updated Q-value for the state-action pair, $Q(s,a)$ is the current Q-value for the same state-action pair, $\alpha$ is the learning rate\footnote{When referring to RL algorithms, a `learning rate' is a hyperparameter used to regulate the rate of updates within the algorithm. This should not be confused with the term `learning ratio' used in this paper, which is used to refer to the ratio of a simulated bidder's bid amount to the item's value.} which determines how much new information overrides old information, and $r$ is the immediate reward received after taking action $a$ in state $s$. To balance the influence of immediate versus future rewards, a discount factor $\gamma$ is used, and the largest Q-value for the next state $s^\prime$ ($\forall a \in s$) is given by $\max_a\left(Q(s',a') - Q(s,a)\right)$.

A key term in Bellman equation is $r + \gamma \max_a\left(Q(s',a') - Q(s,a)\right)$, which represents the difference between the predicted Q-values of the current state-action pair $Q(s,a)$, and the sum of the immediate reward $r$ combined with the discounted value of the `best' action in the next state $Q(s',a')$. This difference, known as the \emph{temporal difference} is added to the existing Q-value in the Q-table (derived in a previous iteration) during the update process. The resulting adjusted Q-value promotes a more accurate estimation of expected future rewards.

Q-learning also requires a search strategy that facilitates the exploration of new actions while exploiting the knowledge gained through previous actions. Typically, this takes the form of an action-selection strategy, whereby a random action, or the `best' action (for a given state), is taken with some probability. Taking a random action encourages exploration by evaluating new action-state pairs and possibly discovering improved policies. Taking the `best' known action aims to maximise the expected reward \citep{tijsma2016comparing}. In this paper, the search strategy employed implements a decaying $\epsilon - greedy$ mechanism where the exploration rate $\epsilon$ decreases over time:

\begin{equation}
    a_t = \begin{cases}
        \arg\max_a Q(s, a) & \text{with probability } 1 - \epsilon_t \\
        \text{random action} & \text{with probability } \epsilon_t,
    \end{cases}
\end{equation}
where $Q(s, a)$ is the Q-value function that estimates the expected future reward for taking action $a$ in state $s$, $t$ is the current time step, and $\arg\max_a Q(s, a)$ represents the action that maximises the Q-value function in the current state $s$. The exploration rate at time $t$ is determined by $\epsilon_t = \epsilon_{\text{max}} \cdot \text{decay\_rate}^t$, where $\epsilon_{\text{max}}$ is the maximum exploration rate (i.e., the initial value of $\epsilon$), $\text{decay\_rate}$ is a constant $[0,1]$ that determines the rate of decay, and $t$ is the current time step.

While the scenarios presented in this paper (i.e., auctions with a single bidding round and one-step episodes) differ from traditional RL problems (with numerous steps per episode), the use of Q-Learning and the Bellman equation remains valid. Although the traditional interpretation of the Bellman equation involves considering future rewards and actions over multiple steps, in a single-step scenario, the equation simplifies to focus on the immediate reward and action taken.

In a one-step episode, the Bellman equation evaluates the immediate reward for a specific action in a given state. It calculates the value of taking that action by considering both the immediate reward and any potential future rewards that the action might influence. Although the concept of future rewards and optimal actions over multiple steps may not be directly applicable in a one-step scenario, evaluating actions based on their immediate consequences is still consistent with the principles of the Bellman equation.

The Q-Learning algorithm faces challenges in multi-unit auction simulation due to large search spaces and non-deterministic and dynamic environments. Storing and processing Q-table can be computationally intractable. Convergence towards an optimal policy is slow, and exploration methods become less effective. The non-deterministic environment leads to Q-value instability and divergence, and the limited state information of agents compounds the problem.

\subsubsection{Deep Q-Learning Network}
A Deep Q-learning network (DQN) combines the principles of Q-learning with the capabilities of Deep Neural Networks (DNN) \citep{mnih2015human}. It addresses the limitations of traditional Q-learning, especially in environments with large or continuous state spaces where maintaining a Q-table becomes impractical. A DNN refers to a subset of machine learning algorithms that use multi-layered structures of nodes or neurons to simulate the decision-making capabilities of the human brain. The use of multiple layers in the network allows for the processing of complex, high-dimensional data. 

In place of a Q-table, DQN uses DNNs as function approximators that estimate the Q-values for each state-action pair. The Q-value function is initialised with a primary neural network and a target network. It selects actions using an $\epsilon\text{-}greedy$ policy and stores experience tuples in a replay memory buffer. The Q-network is updated periodically to minimize the loss between predicted and target Q-values. Through repeated interaction and learning, the Q-network gradually converges to the true Q-values, resulting in an optimal policy that maximizes expected returns.

\subsection{Policy Gradient}

Known as `Vanilla' Policy Gradient (VPG) or `REINFORCE', this RL algorithm optimises policies directly rather than using a value function as Q-Learning does. Policy gradient methods aim to learn a policy $\pi: S \rightarrow A$ that maps states to actions. It is optimised by adjusting policy parameters in the direction that increases the expected return. This is achieved through gradient ascent on the expected return, with the policy represented as a probability distribution over actions. 

The algorithm uses a loss function that guides the update of policy weights based on the expected return function $J(\theta)$. The loss function involves the log probabilities of actions taken in each state multiplied by the discounted rewards, averaged over time steps \citep{sutton2018reinforcement}. The REINFORCE algorithm is synonymous with Policy Gradient approaches and is typically expressed as the following gradient update rule \citep{williams1992simple}:
\begin{equation}
    \theta \leftarrow \theta + \alpha \gamma^t r_t \nabla_\theta \ln \pi_\theta(a_t \vert s_t),
\end{equation}
where $\theta$ represents the parameters or the policy $\pi$, $\alpha$ is the learning rate, $\gamma$ the discount factor, $r_t$ the return from time step $t$, $a_t$ is the action taken at time $t$, and $s_t$ is the state at time $t$. The term $\nabla_\theta \ln \pi_\theta(a_t \vert s_t)$ represents the gradient of the logarithm of the policy's probability of taking action $a_t$ in state $s_t$, with respect to the policy parameters $\theta$.

VPG and REINFORCE both suffer from high variance in gradient estimates, which delays convergence and negatively impacts exploration of the search space. Generally, the principal source of variability in policy gradient approaches stems from the variability of accumulated rewards over long multi-step episodes. While the RL scenarios in this paper comprise single-step episodes, the non-deterministic and dynamic environments used ensure the problem of high variance continues to degrade results.

\subsubsection{Deep Policy Gradient}

Similar to how DQN extends traditional Q-Learning through the use of DNNs, deep policy gradient methods replace the parameterized function of VPG with DNN function approximations. The increased representation power of DNNs allows for efficient search of larger state and action spaces and facilitates the use of techniques to reduce variance and encourage convergence. 

Deep Policy Gradient approaches employ a DNN using the current environment state as an input, and produces either a probability distribution over actions as an output (for discrete action spaces), or parameters of a distribution from which actions are sampled (for continuous action spaces). The DNN's weights and biases are the policy parameters that are then iteratively adjusted during training. Sophisticated optimisers can also be used with Deep Policy methods, which handle noisy gradient estimates much better than VPG approaches. An entropy term is commonly used with the loss function to assist convergence and avoid sub-optimal policies.

While an improvement over VPG, Deep Policy Gradient is very sensitive to the hyperparameter selection. The choice of learning rate, discount factor, and neural network architecture can result in considerable result variation. Unfortunately, choosing the settings is a non-trivial task requiring repeated experimentation and adjustment.

\subsection{Actor-Critic}
The \emph{Actor-Critic} approach combines aspects of both policy-based (the ``Actor'') and value-based (the ``Critic'') learning. The actor component is responsible for selecting actions based on the current policy, that is, a mapping from states to actions. The role of the Critic is to assess the actions taken by the Actor through value function estimation that measures the expected return from a given state using the current policy. The critic's evaluation is then used to update the actor's policy towards more rewarding actions. One key advantage of actor-critic methods is their ability to handle continuous and discrete action spaces. In addition to this versatility, separating the policy and value function estimates can stabilize the learning process when compared to methods using only policy or value function estimation.

\subsubsection{Advantage Actor-Critic}
The \emph{Advantage Actor-Critic} (A2C) method is a variant of the Actor-Critic approach that improves the actor's policy by using the \emph{advantage function}. The advantage function measures the relative quality of an action in a given state compared to the average action in that state. The A2C algorithm addresses some of the key issues with both policy gradient and Q-learning-based methods. Specifically, in using the advantage function, A2C reduces the variance in policy updates, leading to more stable learning; both continuous and discrete action spaces can be handled efficiently (unlike Q-learning approaches); and the method scales well in complex environments due to the separate handling of policy and value function estimations.

The operation of the A2C algorithm proceeds as follows:
\begin{itemize}
    \item During the interaction with the environment phase, the actor assesses the current state of the environment and decides on an action according to its policy. This chosen action is then implemented within the environment, which in turn generates a new state and awards a reward.
    \item The critic appraises the selected action by determining the value of the current and subsequent states. The advantage is then computed by taking the difference between the actual reward received and the sum of the discounted value of the forthcoming state and the value of the present state.
    \item The actor modifies its policy parameters by utilizing gradients obtained from the advantage function, thereby favouring actions that yield positive advantages. At the same time, the critic adjusts its value function parameters with the aim of reducing the discrepancy between the estimated values and the actual returns.
\end{itemize}

An advantage function, $A(s, a)$, encapsulates the difference between the action-value function, $Q(s, a)$, and the state-value function, $V(s)$:
\begin{equation}
    A(s, a) = Q(s, a) - V(s).
\end{equation}

The policy gradient update rule in A2C can be represented as follows, where $\theta$ represents the parameters of the policy network, $\alpha$ is the learning rate, and $\nabla_\theta \log \pi_\theta(a|s)$ is the gradient of the log-probability of taking action $a$ in state $s$ under policy $\pi$ parameterized by $\theta$:
\begin{equation}
\Delta\theta = \alpha \sum_{t=0}^{T} \nabla_\theta \log \pi_\theta(a_t|s_t) A(s_t, a_t).
\end{equation}

The value function (critic) is updated to minimize the difference between the estimated value and the actual return, which can be represented using a loss function, $L(w)$, where $w$ represents the parameters of the value function:

\begin{equation}
L(w) = \frac{1}{2} \sum_{t=0}^{T} \left( R_t - V_w(s_t) \right)^2.
\end{equation}

Here, $R_t$ is the return (cumulative discounted reward) from time $t$.

A2C excels in scenarios where future rewards must be estimated, and decisions must consider long-term outcomes. Consequently, the advantage function plays a crucial role in A2C methods, facilitating the balance between immediate and future rewards. In the single-bid auction scenarios examined in this paper, such features are underutilised and possibly unnecessary. Each auction episode is effectively a single decision with an immediate reward, and the learning process focuses primarily on associating states directly with their resulting rewards without considering future states. Nonetheless, experimental results using A2C (see section \ref{section:Example}) reveal a substantial improvement over simpler policy gradient and Q-learning methods, especially regarding learning stability.

\begin{table}[ht!]
\caption{Principal advantages and disadvantages of the six reinforcement learning algorithms considered in the paper.\label{tbl:Comparison}}
\centering
\small
\include{tables/comparison}
\end{table}

\subsubsection{Proximal Policy Optimisation}

Proximal Policy Optimization (PPO) was introduced in 2017 by John Schulman and colleagues in their paper 'Proximal Policy Optimization Algorithms' \citep{schulman2017proximal}. This paper presented PPO as a simpler alternative to Trust Region Policy Optimization (TRPO) while achieving similar or better performance. PPO works by limiting the change in policy during each update to avoid disruptively large policy updates.

In both RL in general and PPO in particular, a policy refers to an agent's strategy for making decisions. It defines the mapping from perceived environmental states to actions to be taken. In the case of PPO, the algorithm trains a stochastic policy using an on-policy method, whereby exploration is achieved by sampling actions according to the latest version of its stochastic policy.

The policy in RL can be deterministic or stochastic. A deterministic policy directly maps states to actions, while a stochastic policy outputs a probability distribution over actions for a given state. PPO focuses on training a stochastic policy, enhancing exploration and robustness when applied to complex environments. The objective of the PPO algorithm is to improve the policy in a stable and sample-efficient manner, enabling the agent to make better decisions over time.

While PPO and A2C are both state-of-the-art Actor-Critic variants, PPO has several advantages over A2C, particularly in terms of stability, sample efficiency, and flexibility \citep{cheng2021experimental,corecco2023proximal}. Specifically, such advantages include:
\begin{itemize}
    \item Stability: PPO's clipped surrogate objective function (equation \ref{eqn:ppo_clip}) prevents large updates to the policy, ensuring smoother convergence and reducing the likelihood of training instability.
    \item Simplicity: complex hyperparameter tuning and sophisticated optimization techniques are unnecessary.
    \item Sample Efficiency: can learn effective policies from fewer interactions with the environment.
    \item Robustness and Scalability: Handle discrete and continuous action spaces, multiple agents, and parallel environments.
    \item Balanced sample and computational complexity: lower variance and no expensive second-order optimization.
    \item Exploration-Exploitation trade-off: uses the same policy for exploration and exploitation along with entropy regularisation to encourage the former when needed.
    \item Empirical Performance: Empirical results have shown that PPO outperforms A2C in various benchmarks and applications \cite{huang2022a2c}.
\end{itemize}

The key idea behind PPO is to limit the degree of policy change during updates to improve training stability. This is achieved by constraining the policy updates to a `trust region', which prevents the new policy from deviating too far from the old policy. In an auction environment with private information and non-deterministic outcomes, minimizing the size of policy updates is an important feature. Unpredictable by other auction participants (agent or human) during the training of an RL agent can severely compromise learning by misleading an algorithm's search for the best decision. PPO achieves this via a clipped surrogate objective function. The objective of this function is defined as:

\begin{equation}\label{eqn:ppo_clip}
L^{CLIP}(\theta) = \hat{\mathbb{E}}_t \left[ \min (r_t(\theta) \hat{A}_t, \text{clip} (r_t(\theta), 1 - \epsilon, 1 + \epsilon) \hat{A}_t) \right]
\end{equation}
where $L^{CLIP}(\theta)$ is the clipped surrogate objective, 
$\hat{\mathbb{E}}_t$ is the empirical expectation over a batch of training data,  $r_t(\theta)$ is the ratio of the probability of taking an action under the new and old policies, $ \hat{A}_t $ is the estimated advantage of taking an action at time step $t$, and $\text{clip} (x, a, b)$ clips the value of $x$ to be between $a$ and $b$.

The clipped surrogate objective limits the policy update to a range determined by $1 - \epsilon$ and $1 + \epsilon$, where $ \epsilon$ is a hyperparameter that controls the size of the trust region. This ensures that the policy update is not too large, which improves training stability.

Like A2C, using PPO in single-bid auctions does not fully exploit the algorithm's potential. PPO is designed for environments where actions have long-term consequences over multiple steps; however, as shown in Section \ref{section:Example} below, PPO performs exceptionally well in all auction types examined in this paper.

In summary, Table \ref{tbl:Comparison} provides an account of the advantages and disadvantages of each RL algorithm used in this paper. The following section then presents the results from a series of empirical experiments with each of the six algorithms performing as bidders in the three different auction types: Discretionary price, generalized second price, and uniform price.  

\FloatBarrier

\section{Simulations \label{section:Example}}

In this section, we conduct various simulations using the six algorithms described above. Since the primary aim of this paper is to assess various RL algorithms that are suitable for simulating bidding behavior, we maintain a consistent number of bidders across all simulations and only vary the total number of units available to allow for a tractable analysis. We assume that each auction features six bidders, each demanding two units. We explore three scenarios with respect to the total units available for each auction: one with four total units, one with six units, and one with eight units.\footnote{These numbers are chosen to guarantees that we account for all possible equilibria \citep{engelbrecht1998multi, khezr2017new}.} The values of each agent for both units are the same and are random draws from a uniform distribution on $[0,10]$.

The main variable used to assess the learning process in each simulation is called the `learning ratio', which is calculated as the value minus the bid divided by the value. Intuitively, one expects an agent following an optimal strategy to submit bids that proportionally differ from their value in a stable manner. Therefore, the learning ratio is a good metric to test this performance. Once we confirm that the learning patterns demonstrate reasonably consistent learning, we then use revenue and efficiency as the metrics to evaluate auction performance. Revenue is simply the sum of payments made by all winning bidders, and efficiency is calculated as the sum of the $K$ highest values (irrespective of the bids), where $K$ is the number of units available minus the sum of the values of those who won an item. When this term is zero, the auction has allocated the objects in a fully efficient manner.

\subsection{Implementation}

All simulations were programmed using Python 3.10 combined with the ``Gymnasium'' library (for developing and comparing reinforcement learning algorithms)\cite{towers_gymnasium_2023}, and ``Stable Baselines 3'' (for the deep learning and actor-critic algorithm implementations) \cite{stable-baselines3}. All algorithms are implemented with default settings, and no hyper-parameter tuning was conducted. 

Custom Gymnasium environments were created for the simulated auctions, and algorithms from Stable Baselines were modified for use in multi-agent auction scenarios. All simulations were performed on the same computer, running the Linux-based Ubuntu 22.04 operating system with an Intel Core i9-9900KS 4.00GHz CPU, 64Gb RAM, and a Nvidia RTX 3080 GPU with 10Gb RAM (all deep-learning algorithms were run using the GPU).

\subsection{Preliminary simulations}

To compare the performance of each algorithm, we aim to run them in an environment where each method competes against the others. Prior to this, we pre-train each algorithm individually against agents with a random bidding strategy. The objective of this initial training is to facilitate each agent's exploration of the search space and learning the basic mechanics of the auction environments.  In addition, training against agents that make random decisions helps prevent overfitting and co-adaption to other agent's bidding strategies \citep{zhu2024asurv, Zang2023SequentialCM}.

The six RL approaches were initially trained against five agents who made random bids at or below their item values. A total of 54 training sessions were performed with the resultant model (or Q-table) saved. For the six algorithms, training runs were performed using 100,000 episodes for each combination of auction type (DP, GSP and UP) and number of items on offer (4, 6 and 8).

\subsection{Simulation results}

In this subsection, we report the simulation results. We begin by examining the learning ratio for each auction and discuss variations in learning and bidding strategies as the number of available items changes. We then analyze the payoffs that various algorithms achieved in each auction, using these as measures of each algorithm's bidding performance. Subsequently, we focus on the revenue and efficiency of each auction to assess overall performance.

\subsubsection{Learning}
\begin{figure*}[!ht]
    \centering
    \includegraphics[trim=0.3cm 1cm 0.38cm 0.7cm, clip, width=\textwidth]{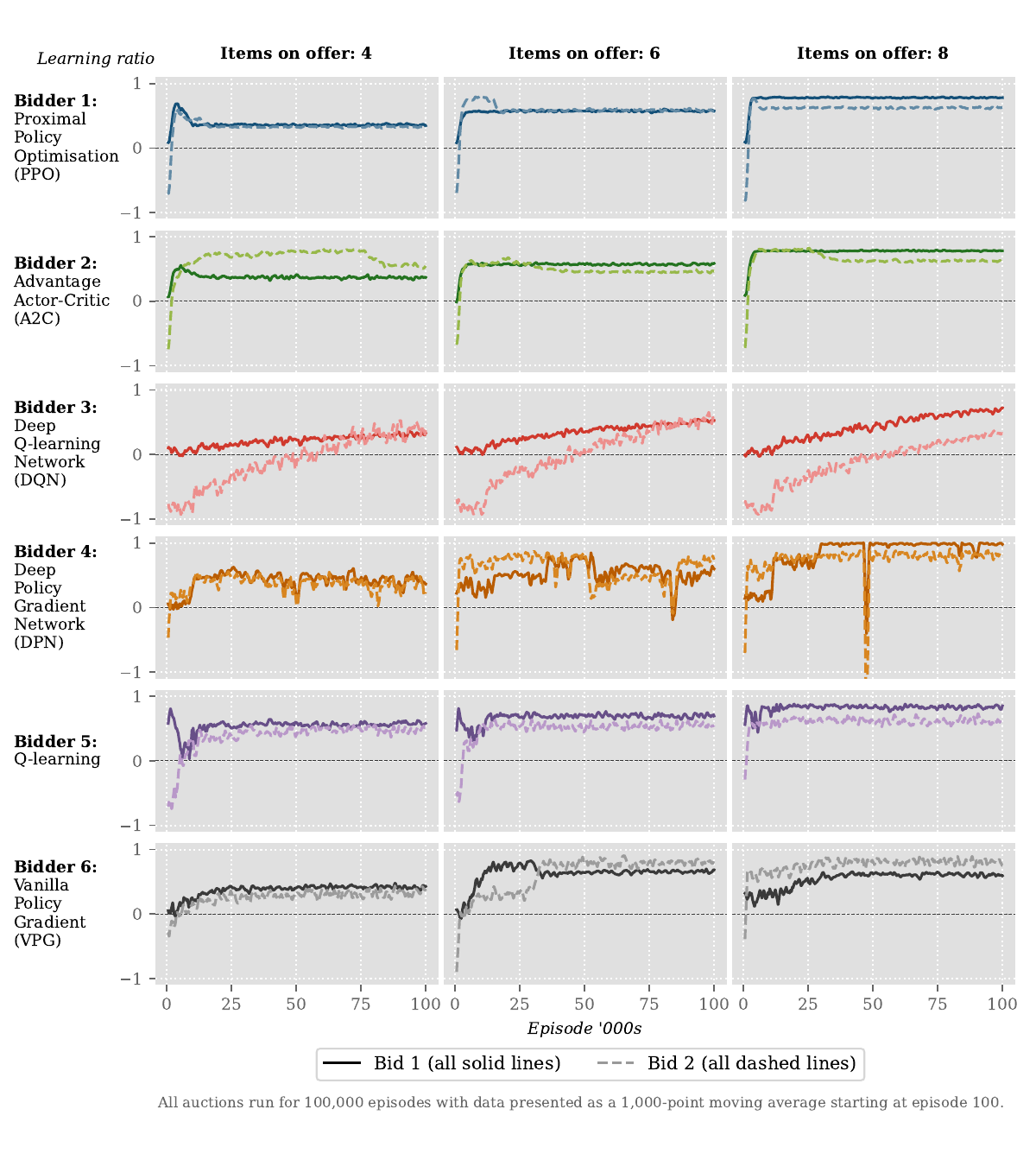}   
    \caption{Bidder learning ratios for the \underline{Discriminatory Price (DP)} auction comparing four, six and eight items on offer with six bidders, each demanding two items. \label{fig:LearningDP}}
\end{figure*}

We begin with the learning results for the three auctions in three different scenarios with respect to the number of items available. Figure \ref{fig:LearningDP} shows the learning ratios of the six algorithms for the discriminatory price auction in three scenarios. As shown in this figure, PPO is the most stable algorithm in terms of the learning ratio and is, in some sense, the fastest. However, all six algorithms demonstrate reasonable convergence to similar strategies by the final episodes. It is important to note that since bids are ranked from highest to lowest, the order in which an agent submits their highest bids (first or second) does not impact the outcome. In all cases, agents submit bids below their values, a straightforward conclusion from the theory, that is, they do not play a dominated strategy \citep{Krishna2009}. Moreover, when more units are available, agents tend to reduce their second bids further, which is also very intuitive.

\begin{figure*}[!ht]
    \centering
    \includegraphics[trim=0.3cm 1cm 0.38cm 0.7cm, clip, width=\textwidth]{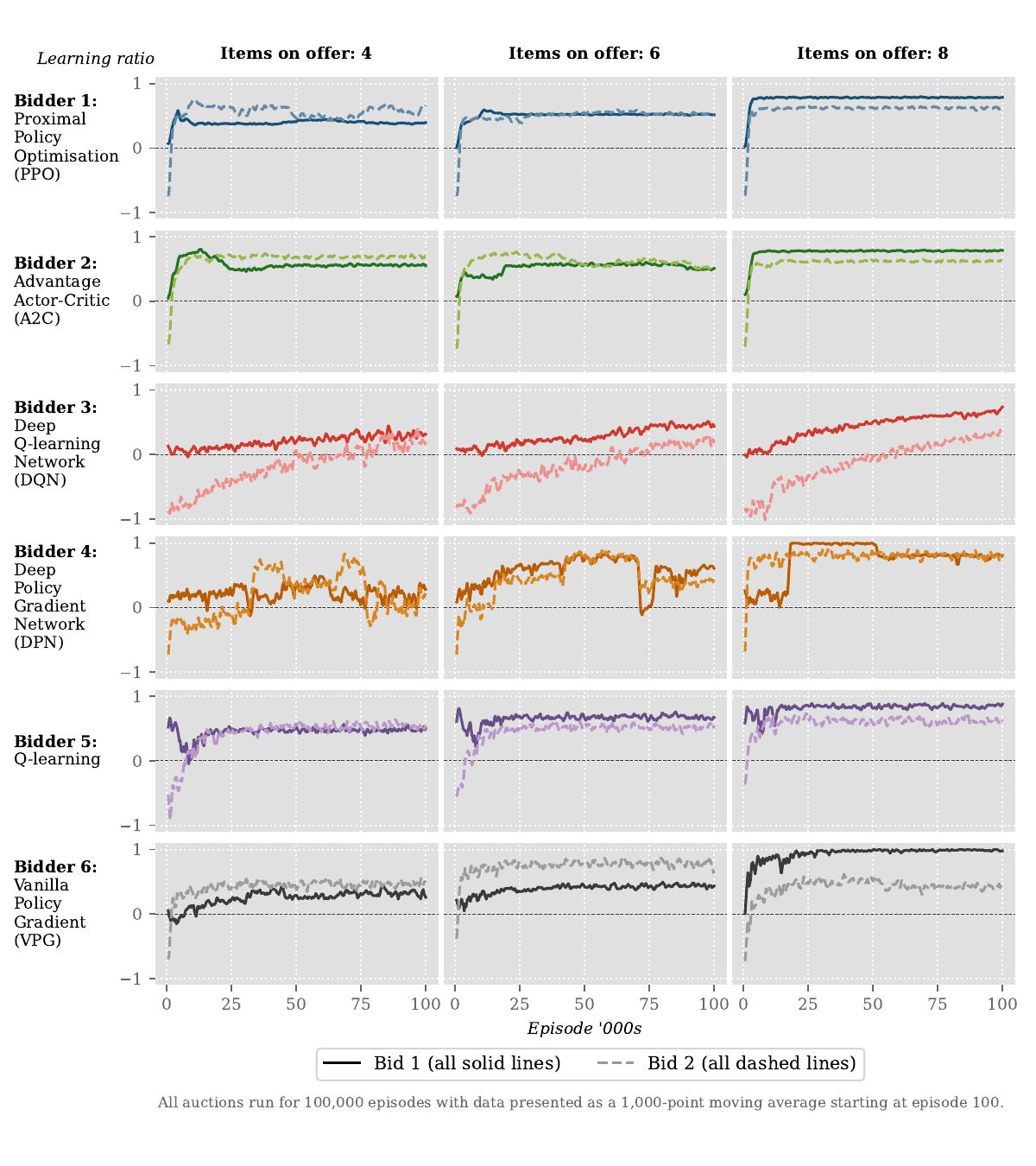}   
    \caption{Bidder learning ratios for the \underline{Generalised Second-Price (GSP)} auction comparing four, six and eight items on offer with six bidders, each demanding two items. \label{fig:LearningGSP}}
\end{figure*}

A similar pattern of learning is observed in Figure \ref{fig:LearningGSP} for the GSP auction. In all cases, agents bid below their values, which supports that truthful bidding is not optimal in GSP \citep{edelman2007internet}. One interesting observation is that the PPO agent submits bids that are very similar to the ones submitted by the same agent in the DP auction. We later show that these bidding strategies resulted in PPO becoming the best performing algorithm in both auctions. The DPN algorithm struggles to learn an optimal strategy and as the number of items increases, its performance worsens. This is mainly because DPN is significantly influenced by stochastic outcomes, and in the GSP format, where the payoff is determined by someone else's bid, DPN struggles to find a consistent learning path. This issue is most pronounced in the Uniform Price auction, where computing an expectation for the highest losing bid from a probabilistic perspective is much more challenging.

\begin{figure*}[!ht]
    \centering
    \includegraphics[trim=0.3cm 1cm 0.38cm 0.7cm, clip, width=\textwidth]{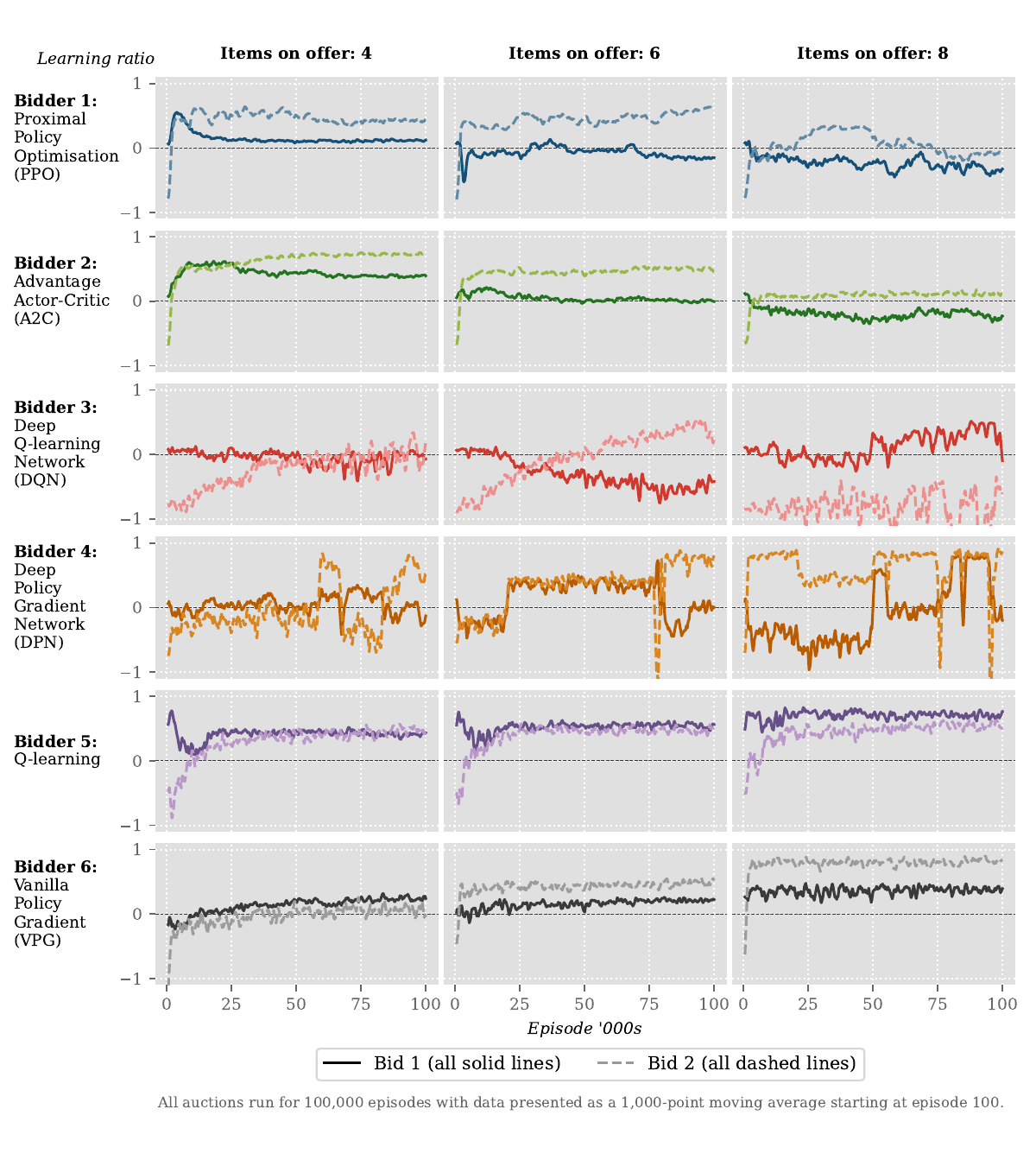}   
    \caption{Bidder learning ratios for the \underline{Uniform-Price (UP)} auction comparing four, six and eight items on offer with six bidders, each demanding two items. \label{fig:LearningUP}}
\end{figure*}

Finally, Figure \ref{fig:LearningUP} illustrates the learning patterns for the uniform price auction. When the number of items is at its lowest, PPO and A2C submit bids that are less truthful, with the level of truthful bidding increasing as the number of items increases to six. However, we observe overbidding with both PPO and A2C when there are eight items available. While this may seem unintuitive, as bidding equal to the value is a weakly dominant strategy in the UP auction \citep{Krishna2009}, the likelihood of incurring a loss from overbidding declines as the number of items increases and could potentially become very low. Thus, the strategies that PPO and A2C agents follow would not necessarily result in lower payoffs compared to other agents. In fact, we later show that the strategies that PPO and A2C followed in UP allowed these two algorithms to achieve the highest payoffs among all agents. Moreover, Q-learning maintains an underbidding strategy across all cases, even increasing the extent of underbidding as the number of items increases while VPG starts with almost truthful bids for both units and underbids more when the total number of items increase. Overall, we observed the highest level of variation in strategies for the UP auction, which shows that despite its simple structure, optimal bidding strategies in this auction are not easy to learn.

\subsubsection{Payoffs}

Having shown the learning patterns of all agents, we next demonstrate their performance in terms of payoffs, which is the most important metric for their behavior. As illustrated in Tables \ref{tbl:BidderResultsItems4}, \ref{tbl:BidderResultsItems6}, and \ref{tbl:BidderResultsItems8}, PPO consistently achieves the highest payoff among the six algorithms without any exceptions. A2C is the second-best performing algorithm in almost all cases, except for two instances where VPG ranks second. The other algorithms vary in terms of their payoff rankings across the three different scenarios and auctions. In particular, Q-learning, despite its simplistic structure, is among the top three algorithms in several scenarios, including DP and GSP with four units and DP with eight units.

\subsubsection{Auction performance}

\begin{figure*}[!ht]
    \centering
    \includegraphics[trim=0cm 0cm 0cm 0cm, clip, width=\textwidth]{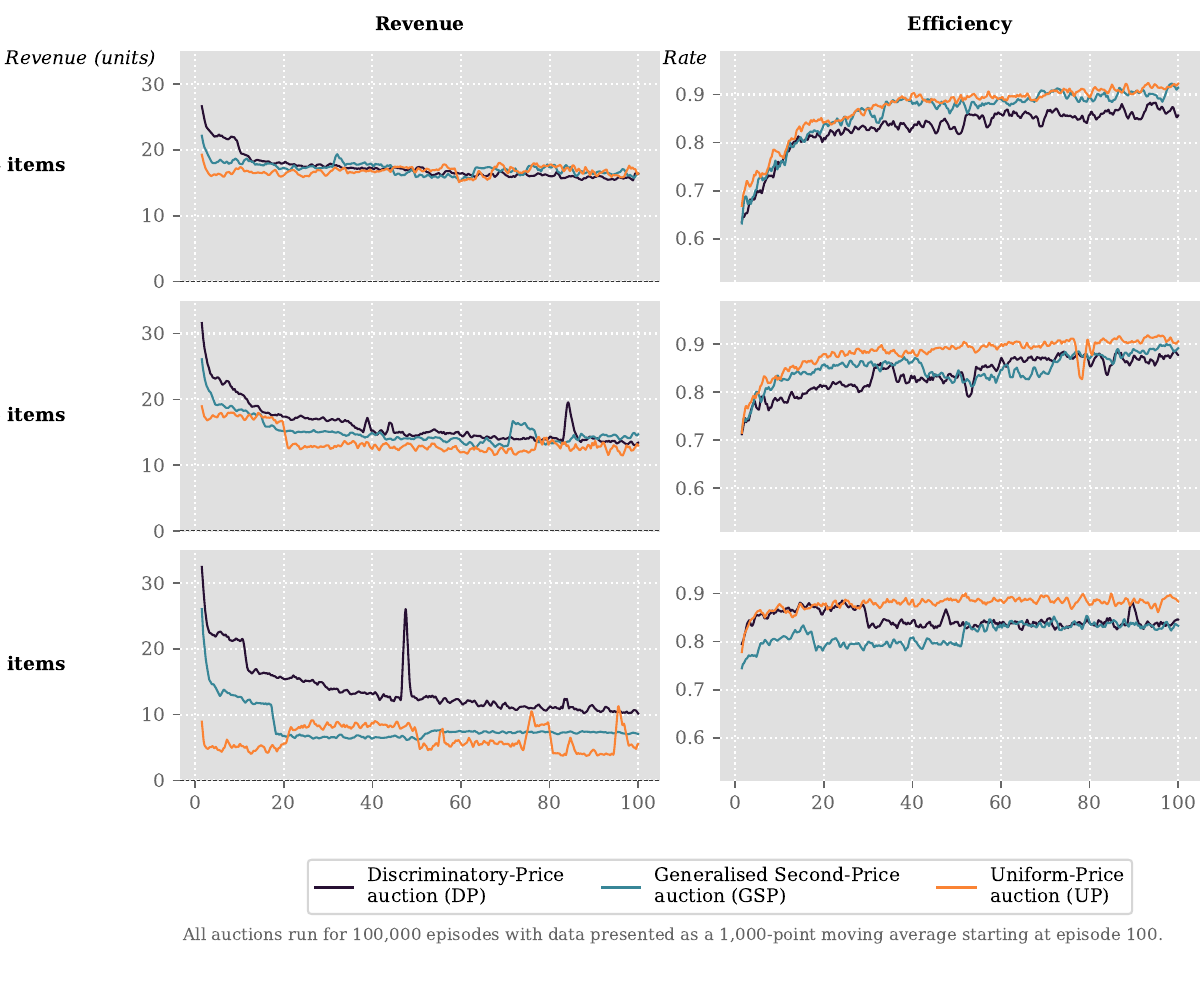}   
    \caption{Comparison of auction performance metrics for the Discriminatory Price (DP), Generalised Second Price (GSP), and Uniform Price (UP) auctions with four, six and eight items on offer. Each auction comprised six bidders (six different reinforcement learning algorithms), each demanding two items. \label{fig:Auctions}}
\end{figure*}

Next, we show how each auction performed in terms of revenue and efficiency. Figure \ref{fig:Auctions} depicts the revenue and efficiency for each auction across the three scenarios. Focusing on later episodes, where the learning patterns stabilize, we observe that with four items available, all three auctions generate similar revenues, with the UP auction slightly outperforming the others. However, as the number of units increases, this ranking changes; with eight items available, the DP auction generates the highest revenue, followed by the GSP and then UP. In terms of efficiency, UP dominates the other two auctions in all scenarios, followed by GSP and DP.

\subsubsection{Simulation results with all PPO agents}

\begin{figure*}[!ht]
    \centering
    \includegraphics[trim=0cm 0cm 0cm 0cm, clip, width=\textwidth]{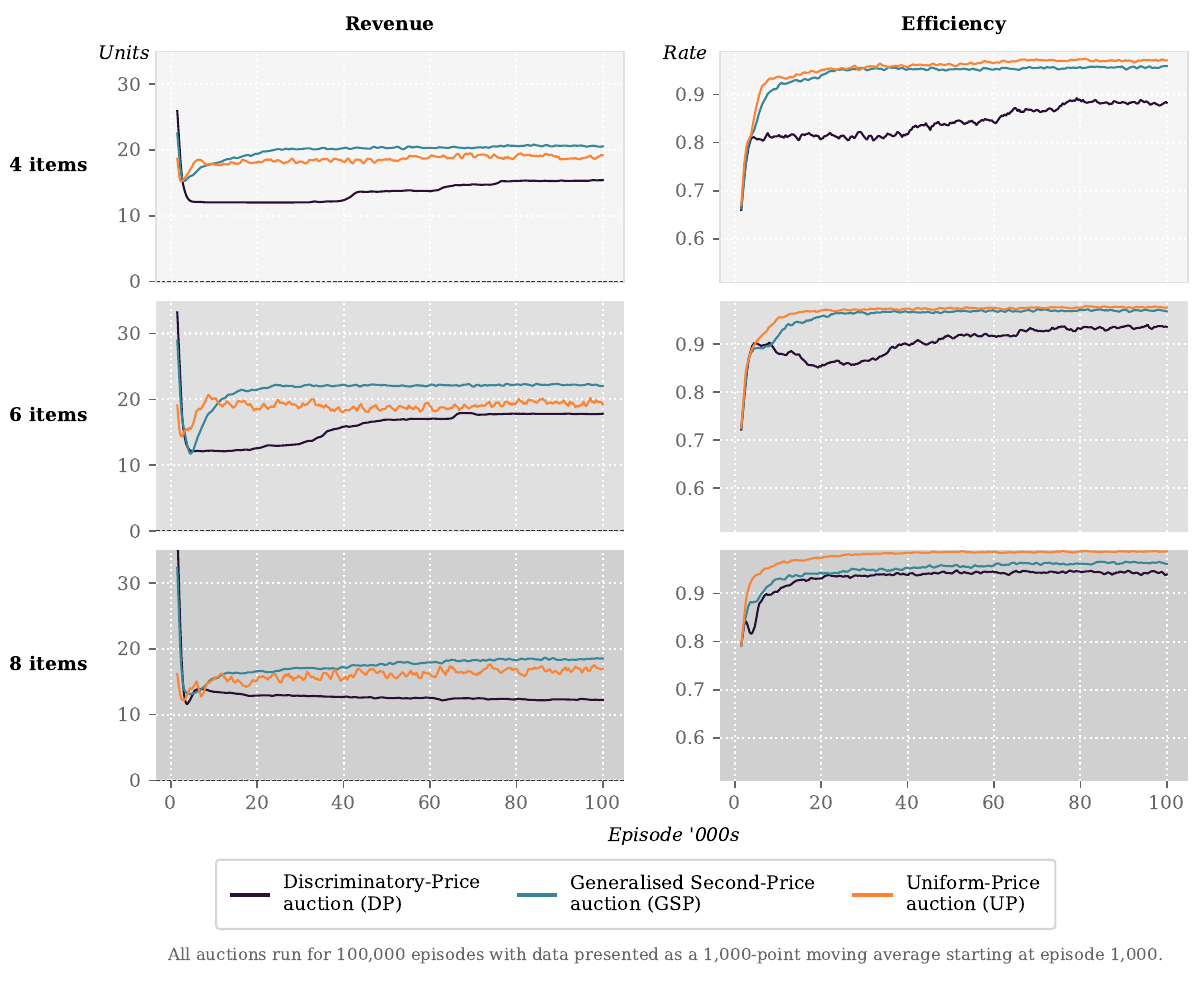}   
    \caption{PPO: Comparison of auction performance metrics for the Discriminatory Price (DP), Generalised Second Price (GSP), and Uniform Price (UP) auctions with four, six and eight items on offer. Each auction comprised six bidders (six PPO algorithms), each demanding two items. \label{fig:AuctionsPPO}}
\end{figure*}

Given the results above, it is clear that the PPO agent significantly outperforms the others in terms of learning to bid optimally. However, observations of the learning patterns indicate that the auction ranking in terms of revenue and efficiency is influenced by other sub-optimal learning behaviors. It is worthwhile to explore how each auction performs when all six agents are PPO. Figure \ref{fig:AuctionsPPO} shows the results regarding the revenue and efficiency rankings of the three auctions when all agents are PPO. The UP auction remains the most efficient among the three, closely followed by GSP. However, the revenue rankings are reversed, with GSP now generating the highest revenue, followed by UP.

Two clear conclusions can be drawn from these observations: first, the uniform price auction consistently performs well in terms of efficiency; second, GSP is the most stable auction in terms of both revenue and efficiency, regardless of whether the agents bid optimally or not.

\section{Conclusions}

In this article, we introduced six reinforcement learning algorithms to simulate bidding behavior in three different multi-unit auctions. While this paper extensively investigates RL-based approaches for simulating auctions, the task is far from complete. There is a pressing need for future studies to explore additional modeling assumptions for agents, including the possibility of allowing more than two bids per agent and increasing the number of items available. We note that such expansions require significant computational resources and time commitment, which were beyond the scope of this research.

\newpage
\section{Appendix: Figures and Tables}

\begin{table}[!h]
\centering
    \caption{Auctions with \underline{four} items on offer: Bidder results by auction type over 100,000 episodes, ranked by total payoff in each auction type. \label{tbl:BidderResultsItems4}}
    \footnotesize
    \def\arraystretch{1.1}
    \setlength{\tabcolsep}{1.1em}
    \include{tables/BidderResultsItems4}
\end{table}

\begin{table}[!hb]
\centering
    \caption{Auctions with \underline{six} items on offer: Bidder results by auction type over 100,000 episodes, ranked by total payoff in each auction type. \label{tbl:BidderResultsItems6}}
    \footnotesize
    \def\arraystretch{1.1}
    \setlength{\tabcolsep}{1.1em}
    \include{tables/BidderResultsItems6}
\end{table}

\begin{table}[!hb]
\centering
    \caption{Auctions with \underline{eight} items on offer: Bidder results by auction type over 100,000 episodes, ranked by total payoff in each auction type. \label{tbl:BidderResultsItems8}}
    \footnotesize
    \def\arraystretch{1.1}
    \setlength{\tabcolsep}{1.1em}
    \include{tables/BidderResultsItems8}
\end{table}

\begin{table}[!hb]
\centering
    \caption{Auction revenue and efficiency by items on offer and auction type over 100,000 episodes \label{tbl:RevenueEfficiency}}
    \footnotesize
    \def\arraystretch{1.3}
    \setlength{\tabcolsep}{0.55em}
    \include{tables/RevenueEfficiency}
\end{table}

\begin{table}[!hb]
\centering
    \caption{PPO: Auction revenue and efficiency by items on offer and auction type over 100,000 episodes \label{tbl:RevenueEfficiencyPPO}}
    \footnotesize
    \def\arraystretch{1.3}
    \setlength{\tabcolsep}{0.55em}
    \include{tables/PPO_RevenueEfficiency}
\end{table}

\clearpage
\bibliographystyle{econ}
\bibliography{bibliography}
\end{document}

%% file: tables/comparison.tex
\begin{tabular}{>{\raggedright\arraybackslash}p{3cm}>{\raggedright\arraybackslash}p{5.5cm}>{\raggedright\arraybackslash}p{5.5cm} }
\hline
\textbf{Algorithm} & \textbf{Advantages} & \textbf{Disadvantages} \\ \hline
\vspace{2pt}Q-Learning & 
\begin{itemize}[leftmargin=*,itemsep=0pt,topsep=0pt,parsep=0pt]
    \item Simple to implement for discrete action spaces
    \item Finds optimal policy given enough time
\end{itemize} & 
\begin{itemize}[leftmargin=*,itemsep=0pt,topsep=0pt,parsep=0pt]
    \item Struggles with large or continuous state spaces
    \item Requires discretization for continuous actions
\end{itemize} \\ \hline
\vspace{2pt}Deep Q-Learning (DQN) & 
\begin{itemize}[leftmargin=*,itemsep=0pt,topsep=0pt,parsep=0pt]
    \item Handles high-dimensional state spaces using deep neural networks
    \item Off-policy learning
\end{itemize} & 
\begin{itemize}[leftmargin=*,itemsep=0pt,topsep=0pt,parsep=0pt]
    \item Struggles with continuous action spaces
    \item Can be sample inefficient
\end{itemize} \\ \hline
\vspace{2pt}Vanilla Policy Gradient & 
\begin{itemize}[leftmargin=*,itemsep=0pt,topsep=0pt,parsep=0pt]
    \item Directly optimizes the policy
    \item Can handle continuous action spaces
\end{itemize} & 
\begin{itemize}[leftmargin=*,itemsep=0pt,topsep=0pt,parsep=0pt]
    \item High variance in gradient estimates
    \item Can be sample inefficient
\end{itemize} \\ \hline
\vspace{2pt}Deep Policy Gradient & 
\begin{itemize}[leftmargin=*,itemsep=0pt,topsep=0pt,parsep=0pt]
    \item Can solve complex problems with high-dimensional inputs
    \item Flexible with continuous action spaces
\end{itemize} & 
\begin{itemize}[leftmargin=*,itemsep=0pt,topsep=0pt,parsep=0pt]
    \item Requires careful tuning of hyperparameters
    \item High computational cost
\end{itemize} \\ \hline
\vspace{2pt}Advantage Actor-Critic (A2C) & 
\begin{itemize}[leftmargin=*,itemsep=0pt,topsep=0pt,parsep=0pt]
    \item Reduces variance of policy gradient estimates
    \item Can handle continuous actions
    \item Balances bias and variance
\end{itemize} & 
\begin{itemize}[leftmargin=*,itemsep=0pt,topsep=0pt,parsep=0pt]
    \item More complex to implement than simpler methods
    \item Requires tuning of advantage estimation
\end{itemize} \\ \hline
\vspace{2pt}Proximal Policy Optimization (PPO) & 
\begin{itemize}[leftmargin=*,itemsep=0pt,topsep=0pt,parsep=0pt]
    \item Stable and robust performance
    \item Simplifies implementation compared to other advanced methods
    \item Effective in various environments
\end{itemize} & 
\begin{itemize}[leftmargin=*,itemsep=0pt,topsep=0pt,parsep=0pt]
    \item More complex than basic policy gradients
    \item May require more computational resources for large-scale problems
\end{itemize} \\ \hline
\end{tabular}

%% file: tables/BidderResultsItems4.tex

\begin{tabular}{p{2.2cm}|ccl|rr|r|r}
\toprule
 &  \multicolumn{3}{c|}{\textbf{Bidder}} & \multicolumn{2}{c|}{\textbf{Payoff}} & \multicolumn{1}{c|}{\textbf{Cost}} &  \\
\textbf{Auction type} & \textbf{Rank} & \textbf{ID} & \textbf{Type} & \multicolumn{1}{r}{\textbf{Total}} & \multicolumn{1}{r|}{\textbf{Mean}} & \multicolumn{1}{r|}{\textbf{Mean}} & \multicolumn{1}{r}{\textbf{Items won}} \\
\toprule
\multirow{6}{2.2cm}{Discretionary Price (DP)} & 1 & 1 & PPO & 253,189 & 2.9 & 4.0 & 86,886 \\
 & 2 & 6 & VPG & 141,213 & 2.0 & 5.1 & 69,335 \\
 & 3 & 5 & QL & 137,913 & 2.2 & 4.0 & 63,710 \\
 & 4 & 2 & A2C & 128,023 & 3.0 & 3.1 & 43,038 \\
 & 5 & 4 & DPN & 109,916 & 2.3 & 3.9 & 47,873 \\
 & 6 & 3 & DQN & -9,577 & -0.1 & 5.2 & 89,158 \\
 \midrule
\multirow{6}{2.2cm}{Generalised Second-price (GSP)} & 1 & 1 & PPO & 225,346 & 3.1 & 4.8 & 72,910 \\
 & 2 & 2 & A2C & 174,360 & 3.0 & 5.0 & 57,324 \\
 & 3 & 5 & QL & 154,940 & 2.6 & 4.6 & 60,610 \\
 & 4 & 4 & DPN & 147,031 & 2.2 & 4.6 & 66,691 \\
 & 5 & 6 & VPG & 134,208 & 2.0 & 6.9 & 68,439 \\
 & 6 & 3 & DQN & 38,024 & 0.5 & 5.6 & 74,026 \\
 \midrule
\multirow{6}{2.2cm}{Uniform price(UP)} & 1 & 1 & PPO & 209,210 & 2.9 & 7.7 & 70,926 \\
 & 2 & 6 & VPG & 192,545 & 2.6 & 6.8 & 74,819 \\
 & 3 & 4 & DPN & 177,993 & 2.4 & 5.9 & 75,657 \\
 & 4 & 2 & A2C & 161,215 & 3.0 & 6.9 & 53,947 \\
 & 5 & 5 & QL & 119,337 & 2.7 & 5.4 & 44,518 \\
 & 6 & 3 & DQN & 90,500 & 1.1 & 6.4 & 80,133 \\
 \bottomrule
\end{tabular}


%% file: tables/BidderResultsItems6.tex

\begin{tabular}{p{2.2cm}|ccl|rr|r|r}
\toprule
 &  \multicolumn{3}{c|}{\textbf{Bidder}} & \multicolumn{2}{c|}{\textbf{Payoff}} & \multicolumn{1}{c|}{\textbf{Cost}} &  \\
\textbf{Auction type} & \textbf{Rank} & \textbf{ID} & \textbf{Type} & \multicolumn{1}{r}{\textbf{Total}} & \multicolumn{1}{r|}{\textbf{Mean}} & \multicolumn{1}{r|}{\textbf{Mean}} & \multicolumn{1}{r}{\textbf{Items won}} \\
\toprule
\multirow{6}{2.2cm}{Discretionary Price (DP)} & 1 & 1 & PPO & 453,766 & 3.7 & 2.1 & 123,852 \\
 & 2 & 2 & A2C & 366,916 & 3.9 & 2.0 & 94,967 \\
 & 3 & 6 & VPG & 286,966 & 3.6 & 2.4 & 80,754 \\
 & 4 & 4 & DPN & 269,956 & 3.3 & 2.7 & 82,612 \\
 & 5 & 5 & QL & 253,881 & 2.7 & 2.8 & 92,859 \\
 & 6 & 3 & DQN & 139,945 & 1.1 & 4.0 & 124,956 \\
 \midrule
\multirow{6}{2.2cm}{Generalised Second-price (GSP)} & 1 & 1 & PPO & 463,650 & 3.8 & 2.9 & 122,098 \\
 & 2 & 2 & A2C & 358,361 & 4.0 & 2.3 & 89,536 \\
 & 3 & 6 & VPG & 309,214 & 3.5 & 4.1 & 88,471 \\
 & 4 & 3 & DQN & 290,137 & 2.0 & 4.0 & 142,628 \\
 & 5 & 5 & QL & 262,489 & 3.0 & 3.0 & 88,643 \\
 & 6 & 4 & DPN & 245,347 & 3.6 & 2.7 & 68,624 \\
 \midrule
\multirow{6}{2.2cm}{Uniform price(UP)} & 1 & 1 & PPO & 448,183 & 4.0 & 6.6 & 113,126 \\
 & 2 & 2 & A2C & 437,020 & 3.9 & 5.7 & 112,489 \\
 & 3 & 6 & VPG & 378,679 & 4.0 & 4.9 & 93,767 \\
 & 4 & 3 & DQN & 339,120 & 2.9 & 6.7 & 116,875 \\
 & 5 & 4 & DPN & 329,467 & 3.7 & 4.6 & 90,033 \\
 & 6 & 5 & QL & 285,283 & 3.9 & 4.2 & 73,710\\
\bottomrule
\end{tabular}


%% file: tables/BidderResultsItems8.tex

\begin{tabular}{p{2.2cm}|ccl|rr|r|r}
\toprule
 &  \multicolumn{3}{c|}{\textbf{Bidder}} & \multicolumn{2}{c|}{\textbf{Payoff}} & \multicolumn{1}{c|}{\textbf{Cost}} &  \\
\textbf{Auction type} & \textbf{Rank} & \textbf{ID} & \textbf{Type} & \multicolumn{1}{r}{\textbf{Total}} & \multicolumn{1}{r|}{\textbf{Mean}} & \multicolumn{1}{r|}{\textbf{Mean}} & \multicolumn{1}{r}{\textbf{Items won}} \\
\toprule
\multirow{6}{2.2cm}{Discretionary   Price  (DP)} & 1 & 1 & PPO & 691,272 & 4.1 & 1.0 & 168,251 \\
 & 2 & 2 & A2C & 649,851 & 4.2 & 1.0 & 154,506 \\
 & 3 & 5 & QL & 415,899 & 3.6 & 1.5 & 116,527 \\
 & 4 & 6 & VPG & 357,015 & 3.3 & 2.5 & 107,338 \\
 & 5 & 3 & DQN & 332,151 & 1.7 & 2.8 & 190,720 \\
 & 6 & 4 & DPN & 226,261 & 3.6 & 1.5 & 62,658 \\
\midrule
\multirow{6}{2.2cm}{Generalised   Second-price (GSP)} & 1 & 1 & PPO & 711,596 & 4.3 & 1.0 & 164,451 \\
 & 2 & 2 & A2C & 701,923 & 4.3 & 1.0 & 164,603 \\
 & 3 & 3 & DQN & 576,871 & 3.1 & 2.8 & 188,601 \\
 & 4 & 5 & QL & 450,890 & 4.0 & 1.5 & 111,465 \\
 & 5 & 4 & DPN & 389,694 & 4.6 & 1.4 & 84,582 \\
 & 6 & 6 & VPG & 262,028 & 3.0 & 1.7 & 86,298 \\
 \midrule
\multirow{6}{2.2cm}{Uniform price (UP)} & 1 & 1 & PPO & 725,297 & 4.5 & 5.6 & 160,439 \\
 & 2 & 2 & A2C & 715,136 & 4.6 & 5.5 & 156,397 \\
 & 3 & 3 & DQN & 671,326 & 3.7 & 4.8 & 182,148 \\
 & 4 & 4 & DPN & 547,658 & 4.9 & 5.2 & 111,755 \\
 & 5 & 6 & VPG & 532,158 & 5.3 & 3.0 & 101,297 \\
 & 6 & 5 & QL & 389,310 & 4.4 & 2.9 & 87,964 \\
\bottomrule
\end{tabular}


%% file: tables/RevenueEfficiency.tex

\begin{tabular}{cl|rrrr|rrr}
 \toprule
\multirow{2}{*}{\textbf{Items}} & \multirow{2}{*}{\textbf{Auction type}} & \multicolumn{4}{c|}{\textbf{Revenue}} & \multicolumn{3}{c}{\textbf{Efficiency}} \\
 & & \textbf{Total} & \textbf{Mean} & \textbf{Min} & \textbf{Max} & \textbf{Mean} & \textbf{Min} & \textbf{Max} \\
 \toprule
\multirow{3}{*}{4} & Discretionary Price  (DP)      & 1,738,683 & 17.39 & 8 & 36 & 0.83 & 0.04 & 1.00 \\
                   & Generalised Second-price (GSP) & 1,721,494 & 17.21 & 3 & 36 & 0.86 & 0.04 & 1.00 \\
                   & Uniform price (UP)             & 1,679,288 & 16.79 & 0 & 36 & 0.87 & 0.00 & 1.00 \\
 \midrule
\multirow{3}{*}{6} & Discretionary Price  (DP)      & 1,628,580 & 16.29 & 7 & 53 & 0.84 & 0.15 & 1.00 \\
                   & Generalised Second-price (GSP) & 1,520,078 & 15.2  & 4 & 52 & 0.85 & 0.22 & 1.00 \\
                   & Uniform price (UP)             & 1,366,080 & 13.66 & 0 & 48 & 0.88 & 0.20 & 1.00 \\
 \midrule
\multirow{3}{*}{8} & Discretionary Price  (DP)      & 1,400,719 & 14.01 & 3 & 63 & 0.84 & 0.34 & 1.00 \\
                   & Generalised Second-price (GSP) & 831,480   & 8.31  & 1 & 65 & 0.81 & 0.21 & 1.00 \\
                   & Uniform price (UP)             & 642,584   & 6.43  & 0 & 56 & 0.88 & 0.30 & 1.00 \\
 \bottomrule
\end{tabular}


%% file: tables/PPO_RevenueEfficiency.tex

\begin{tabular}{cl|rrrr|rrr}
 \toprule
\multirow{2}{*}{\textbf{Items}} & \multirow{2}{*}{\textbf{Auction type}} & \multicolumn{4}{c|}{\textbf{Revenue}} & \multicolumn{3}{c}{\textbf{Efficiency}} \\
 & & \textbf{Total} & \textbf{Mean} & \textbf{Min} & \textbf{Max} & \textbf{Mean} & \textbf{Min} & \textbf{Max} \\
 \toprule
\multirow{3}{*}{4} & Discretionary Price  (DP)      & 1,354,492 & 13.68 & 9 & 35 & 0.88 & 0.22 & 1.00 \\
                   & Generalised Second-price (GSP) & 1,970,021 & 19.90 & 5 & 33 & 0.97 & 0.06 & 1.00 \\
                   & Uniform price (UP)             & 1,833,460 & 18.52 & 4 & 32 & 0.98 & 0.35 & 1.00 \\
 \midrule
\multirow{3}{*}{6} & Discretionary Price  (DP)      & 1,567,221 & 15.83 & 7 & 47 & 0.95 & 0.36 & 1.00 \\
                   & Generalised Second-price (GSP) & 2,123,696 & 21.45 & 7 & 42 & 0.99 & 0.32 & 1.00 \\
                   & Uniform price (UP)             & 1,875,216 & 18.94 & 0 & 48 & 0.99 & 0.43 & 1.00 \\     
 \midrule
\multirow{3}{*}{8} & Discretionary Price  (DP)      & 1,267,547 & 12.80 & 6 & 56 & 0.97 & 0.50 & 1.00 \\
                   & Generalised Second-price (GSP) & 1,723,650 & 17.41 & 9 & 56 & 0.98 & 0.50 & 1.00 \\
                   & Uniform price (UP)             & 1,585,248 & 16.01 & 0 & 56 & 0.99 & 0.60 & 1.00\\
 \bottomrule
\end{tabular}


%% file: bibliography.bib
@phdthesis{watkins1989learning,
  title={Learning from Delayed Rewards},
  author={Watkins, Christopher John Cornish Hellaby},
  year={1989},
  school={King's College, Cambridge}
}

@book{bellman1957dynamic,
title={Dynamic Programming},
author={Bellman, Richard},
year={1957},
publisher={Princeton University Press}
}

@INPROCEEDINGS{tijsma2016comparing,
  author={Tijsma, Arryon D. and Drugan, Madalina M. and Wiering, Marco A.},
  booktitle={2016 IEEE Symposium Series on Computational Intelligence (SSCI)}, 
  title={Comparing exploration strategies for Q-learning in random stochastic mazes}, 
  year={2016},
  volume={},
  number={},
  pages={1-8},
  doi={10.1109/SSCI.2016.7849366}}

@article{mnih2015human,
  title={Human-level control through deep reinforcement learning},
  author={Mnih, Volodymyr and Kavukcuoglu, Koray and Silver, David and Rusu, Andrei A. and Veness, Joel and Bellemare, Marc G. and Graves, Alex and Riedmiller, Martin and Fidjeland, Andreas K. and Ostrovski, Georg and others},
  journal={Nature},
  volume={518},
  number={7540},
  pages={529--533},
  year={2015},
  publisher={Nature Publishing Group},
  doi={10.1038/nature14236}
}

@article{williams1992simple,
  author = {Williams, Ronald J.},
  title = {Simple statistical gradient-following algorithms for connectionist reinforcement learning},
  journal = {Machine Learning},
  volume = {8},
  number = {3},
  pages = {229--256},
  year = {1992},
  doi = {10.1007/BF00992696}
}

@book{sutton2018reinforcement,
  title={Reinforcement Learning: An Introduction},
  author={Sutton, Richard S and Barto, Andrew G},
  edition={2},
  year={2018},
  publisher={MIT Press}
}

@inproceedings{cheng2021experimental,
  title={Experimental Evaluation of Proximal Policy Optimization and Advantage Actor-Critic RL Algorithms using MiniGrid Environment},
  author={Cheng, Wen-Chung (Andy) and Ni, Zhen and Zhong, Xiangnan},
  booktitle={34th Florida Conference on Recent Advances in Robotics (FCRAR 2021)},
  year={2021},
  organization={Florida Atlantic University},
  address={Boca Raton, United States},
}

@Article{corecco2023proximal,
AUTHOR = {Corecco, Samuel and Adorni, Giorgia and Gambardella, Luca Maria},
TITLE = {Proximal Policy Optimization-Based Reinforcement Learning and Hybrid Approaches to Explore the Cross Array Task Optimal Solution},
JOURNAL = {Machine Learning and Knowledge Extraction},
VOLUME = {5},
YEAR = {2023},
NUMBER = {4},
PAGES = {1660--1679},
URL = {https://www.mdpi.com/2504-4990/5/4/82},
ISSN = {2504-4990},
DOI = {10.3390/make5040082}
}

@misc{huang2022a2c,
      title={A2C is a special case of PPO}, 
      author={Shengyi Huang and Anssi Kanervisto and Antonin Raffin and Weixun Wang and Santiago Ontañón and Rousslan Fernand Julien Dossa},
      year={2022},
      eprint={2205.09123},
      archivePrefix={arXiv},
      primaryClass={cs.LG}
}

@misc{towers_gymnasium_2023,
        title = {Gymnasium},
        url = {https://zenodo.org/record/8127025},
        urldate = {2023-07-08},
        publisher = {Zenodo},
        author = {Towers, Mark and Terry, Jordan K. and Kwiatkowski, Ariel and Balis, John U. and Cola, Gianluca de and Deleu, Tristan and Goulão, Manuel and Kallinteris, Andreas and KG, Arjun and Krimmel, Markus and Perez-Vicente, Rodrigo and Pierré, Andrea and Schulhoff, Sander and Tai, Jun Jet and Shen, Andrew Tan Jin and Younis, Omar G.},
        month = mar,
        year = {2023},
        doi = {10.5281/zenodo.8127026},
}

@article{stable-baselines3,
  author  = {Antonin Raffin and Ashley Hill and Adam Gleave and Anssi Kanervisto and Maximilian Ernestus and Noah Dormann},
  title   = {Stable-Baselines3: Reliable Reinforcement Learning Implementations},
  journal = {Journal of Machine Learning Research},
  year    = {2021},
  volume  = {22},
  number  = {268},
  pages   = {1-8},
  url     = {http://jmlr.org/papers/v22/20-1364.html}
}

@article{Khezr2022,
  title={A review of multiunit auctions with homogeneous goods},
  author={Khezr, Peyman and Cumpston, Anne},
  journal={Journal of Economic Surveys},
  volume={36},
  number={4},
  pages={1225--1247},
  year={2022},
  publisher={Wiley Online Library}
}

@book{bichler2017handbook,
  title={Handbook of spectrum auction design},
  author={Bichler, Martin and Goeree, Jacob K},
  year={2017},
  publisher={Cambridge University Press}
}

@article{silver2021reward,
  title={Reward is enough},
  author={Silver, David and Singh, Satinder and Precup, Doina and Sutton, Richard S},
  journal={Artificial Intelligence},
  volume={299},
  pages={103535},
  year={2021},
  publisher={Elsevier}
}

@article{vickrey1961counterspeculation,
  title={Counterspeculation, auctions, and competitive sealed tenders},
  author={Vickrey, William},
  journal={The Journal of finance},
  volume={16},
  number={1},
  pages={8--37},
  year={1961},
  publisher={JSTOR}
}

@article{back1993auctions,
  title={Auctions of divisible goods: On the rationale for the treasury experiment},
  author={Back, Kerry and Zender, Jaime F},
  journal={The Review of Financial Studies},
  volume={6},
  number={4},
  pages={733--764},
  year={1993},
  publisher={Oxford University Press}
}

@article{edelman2007internet,
  title={Internet advertising and the generalized second-price auction: Selling billions of dollars worth of keywords},
  author={Edelman, Benjamin and Ostrovsky, Michael and Schwarz, Michael},
  journal={American economic review},
  volume={97},
  number={1},
  pages={242--259},
  year={2007},
  publisher={American Economic Association}
}

@article{ausubel2014demand,
  title={Demand reduction and inefficiency in multi-unit auctions},
  author={Ausubel, Lawrence M and Cramton, Peter and Pycia, Marek and Rostek, Marzena and Weretka, Marek},
  journal={The Review of Economic Studies},
  volume={81},
  number={4},
  pages={1366--1400},
  year={2014},
  publisher={Oxford University Press}
}

@article{roughgarden2017price,
  title={The price of anarchy in auctions},
  author={Roughgarden, Tim and Syrgkanis, Vasilis and Tardos, Eva},
  journal={Journal of Artificial Intelligence Research},
  volume={59},
  pages={59--101},
  year={2017}
}

@inproceedings{morgenstern2016learning,
  title={Learning simple auctions},
  author={Morgenstern, Jamie and Roughgarden, Tim},
  booktitle={Conference on Learning Theory},
  pages={1298--1318},
  year={2016},
  organization={PMLR}
}

@inproceedings{banchio2022artificial,
  title={Artificial intelligence and auction design},
  author={Banchio, Martino and Skrzypacz, Andrzej},
  booktitle={Proceedings of the 23rd ACM Conference on Economics and Computation},
  pages={30--31},
  year={2022}
}

@article{khezr2024strategic,
  title={Strategic Bidding in Knapsack Auctions},
  author={Khezr, Peyman and Mohan, Vijay and Page, Lionel},
  journal={arXiv preprint arXiv:2403.07928},
  year={2024}
}

@article{hailu2007designing,
  title={Designing Multi-unit Multiple Bid Auctions: An Agent-based Computational Model of Uniform, Discriminatory and Generalised Vickrey Auctions},
  author={Hailu, Atakelty and Thoyer, Sophie},
  journal={Economic Record},
  volume={83},
  pages={S57--S72},
  year={2007},
  publisher={Wiley Online Library}
}

@article{schulman2017proximal,
  title={Proximal policy optimization algorithms},
  author={Schulman, John and Wolski, Filip and Dhariwal, Prafulla and Radford, Alec and Klimov, Oleg},
  journal={arXiv preprint arXiv:1707.06347},
  year={2017}
}

@article{engelbrecht1998multi,
  title={Multi-unit auctions with uniform prices},
  author={Engelbrecht-Wiggans, Richard and Kahn, Charles M},
  journal={Economic theory},
  volume={12},
  pages={227--258},
  year={1998},
  publisher={Springer}
}

@article{khezr2017new,
  title={A new characterization of equilibrium in multiple-object uniform-price auctions},
  author={Khezr, Peyman and Menezes, Flavio M},
  journal={Economics letters},
  volume={157},
  pages={53--55},
  year={2017},
  publisher={Elsevier}
}

@book{Krishna2009,
 author = {Vijay Krishna},
 title = {Auction Theory, 2nd Edition},
 publisher = {Academic Press},
 address = {San Diego},
 year = {2009}
 }

@article{zhu2024asurv,
  author    = {Changxi Zhu and Mehdi Dastani and Shihan Wang},
  title     = {A survey of multi-agent deep reinforcement learning with communication},
  journal   = {Autonomous Agents and Multi-Agent Systems},
  volume    = {38},
  number    = {1},
  pages     = {4},
  year      = {2024},
  doi       = {10.1007/s10458-023-09633-6},
}

@inproceedings{Zang2023SequentialCM,
  title={Sequential Cooperative Multi-Agent Reinforcement Learning},
  author={Yifan Zang and Jinmin He and Kai Li and Haobo Fu and Qiang Fu and Junliang Xing},
  booktitle={Adaptive Agents and Multi-Agent Systems},
  year={2023},
  url={https://api.semanticscholar.org/CorpusID:258845637}
}
